# γ-Families with Halos Observed by X-ray Emulsion Chamber in EAS and the Estimate of the p+He Fraction in Primary Cosmic Rays at $E_0$ = 1-100 PeV

R. A. Mukhamedshin[b], V. S. Puchkov[a], S. E. Pyatovsky[a], and S. B. Shaulov[a]

[a] *Lebedev Physical Institute, Russian Academy of Sciences, Leninskii pr. 53, Moscow, 119991 Russia; e-mail: vgsep@ya.ru*

[b] *Institute for Nuclear Research, Russian Academy of Sciences, pr. 60-letiya Octyabrya 7a, Moscow, 117315 Russia*

**Abstract** – The experimental γ-families in which events called "halos" were detected using *X*-ray emulsion chambers (XRECs) of the PAMIR experiment are described. The halo nature is explained within the standard model of nuclear interactions. Based on events with halo, the fraction of p+He in primary cosmic rays (PCR) was estimated. It is shown that the PCR mass composition remains mixed in the range $E_0$ = 1-100 PeV.

The XREC method is widely used to study EAS cores in CR. The schematic diagram of the PAMIR XREC setup is shown in Fig. 1. The high-energy component of EAS cores ($E_\gamma \geq 2$ TeV) is recorded on an *X*-ray film placed between Pb layers as separate dark spots. The high resolution of the *X*-ray film and the high detection threshold of γ-rays allow XRECs to detect narrow collimated beams of high-energy particles as groups of dark spots genetically related to individual EAS and called γ-families.

The use of *X*-ray films with double-sided emulsion layers separated by a plastic substrate ~200 μm thick allows the determination of the coordinates, zenith and azimuth angles of the EAS electron-photon component (EPC) with an accuracy of ~100 μm, 3° and 15°, respectively.

The density on the *X*-ray film is proportional to the number of γ-quanta and $e^\pm$ in the EPC initiated in Pb, i.e., is proportional to the energy $E_\gamma$ of the γ-quantum which caused the electromagnetic cascade in Pb. The energy determination accuracy for photometry is $\sigma(E_\gamma)/E_\gamma$ = 0.2-0.3.

As the primary particle energy increases, individual spots of γ-rays begin to overlap forming areas of increased optical density. Such families are referred to as γ-families with halo.

To analyze γ-families with halo, the MC0 model [1] close to the QGSJ model and well describing characteristics of experimental data obtained using XRECs is used. An analysis of the experimental data of the PAMIR XREC using the MC0 showed that more than 90% of detected γ-families were formed by the p+He PCR component and can be used to estimate the p+He fraction in PCR.

*γ-families with halo.* The events with halo are observed in experimental γ-families beginning with $E_0 \sim$ 3-5 PeV. The event with halo, observed for the first time in the Japanese-Brazilian experiment at the Chakaltaya mountain (Bolivia), was called the "Andromeda". The "Andromeda" halo was attributed to exotic events in nuclear interactions.

In what follows, events with halo were observed in other experiments with XRECs, including the XREC PAMIR. The XREC PAMIR statistics includes 61 γ-families with halo at a total exposure of 3000 m$^2$·year·sr. The largest in area were the "FIANIT" (Fig. 2) and "Tadzhikistan" halos.

To study the halo phenomena, γ-families in the range of $E_0$ = 0.3-3000 PeV were simulated according to the MC0 model with the following criteria:

- the minimum energy of families at the observation level is 100 TeV;
- the minimum number of particles in families is $N_{min}$ = 3;

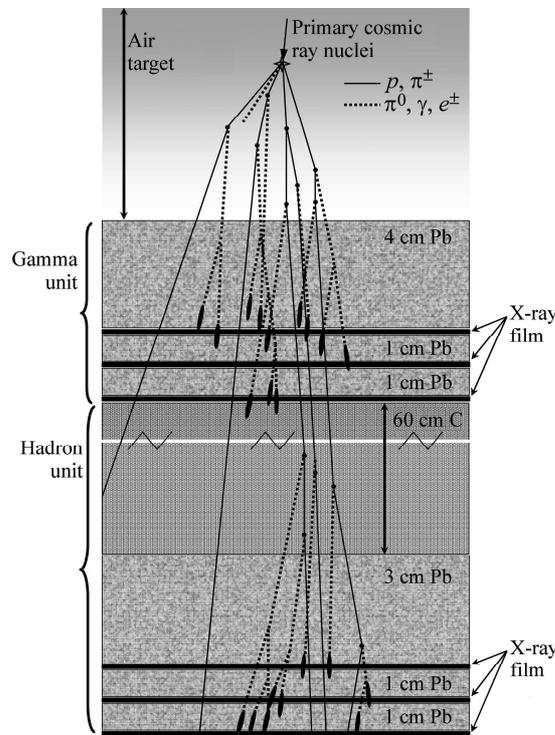
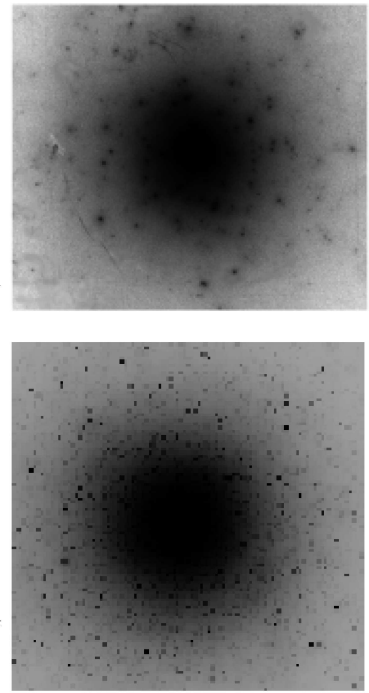

Fig. 1                                      Fig. 2

**Fig. 1.** Schematic diagram of the PAMIR XREC experiment.

**Fig. 2.** "FIANIT" halo scan ($E_0$ estimate by the energy release in the XREC ~400 PeV, $S_{halo}$ = 1020 mm$^2$) (left); the calculated halo formed by primary $p$ ($E_p$ = 420 PeV, $S_{halo}$ = 2100 mm$^2$) (right).

- PCR are presented by particles of 9 groups of nuclei: p, He, C, N, O, Mg, Si, V, Fe;
- the observation level depth is 594 g/cm$^2$ for PAMIR XRECs [2, 3] and 695 g/cm$^2$ for the EAS+XREC of the HADRON experiment (Tien-Shan) [4];
- the energy threshold of 100 GeV for γ-rays and stable hadrons coming to the XREC from the atmosphere;
- the zenith angle of PCR nucleus arrival is in the range from 0 to 0.9 rad.

The halo calculation was performed using γ-families simulated by the Monte Carlo method within the MC0 model using the "chessboard" algorithm. The preliminary selection of γ-families corresponds to PAMIR XREC experimental data processing:

- selection of γ-families;

- γ-families with $\Sigma E_\gamma \geq 100$ TeV ($E_\gamma \geq 4$ TeV) and γ-rays with the distance $R \leq 15$ cm to the energetically weighted center of the γ-family.

In "chessboard" algorithm used in the halo calculation, the spatial distribution functions (SDF) of γ-rays in Pb, which were calculated for the PAMIR XREC configuration, were used. The "chessboard" algorithm SDFs consider the distance of γ-rays to the EAS axis ($R$), the cascade development depth in Pb, the γ-ray energy, and the probability that the first interaction occurs in the XREC depth. SDFs are matched with each other in the axial approximation and the approximation of large $R$ and consider the Landau-Pomeranchuk-Migdal effect.

The selection conditions of γ-families with halo in PAMIR XRECs and for calculations within the MC0 model are

- $\Sigma E_\gamma \geq 500$ TeV;

- the halo area is $S_{D=0.5}$ within an isodense with optical density $D = 0.5$ $S_{D=0.5} \geq 4$ mm² for a halo with one center;

however, in some cases, the halo can contain several centers; in this case, the halo is called the "multicenter" one, and its total area is

$\Sigma S_{iD=0.5} \geq 4$ mm², if $S_{iD=0.5} \geq 1$ mm².

In γ-families with $\Sigma E_\gamma \geq 500$ TeV ($E_\gamma \geq 4$ TeV), high-energy γ-ray cascades in Pb XRECs are overlapped, forming areas with high optical density on the $X$-ray film. The calculations by the "chessboard" algorithm using the SDF made it possible to obtain halos similar to experimental ones. Figure 2 shows the "FIANIT" experimental halo (left) and the calculated halo formed by primary $p$ ($E_p = 420$ PeV) (right) with areas $S_{halo} = 1020$ and $2100$ mm², respectively. As follows from Fig. 2, the formation mechanism of γ-families with halo is well described within the standard model of nuclear interactions.

Calculations showed that γ-families without halo with $\Sigma E_\gamma = 100$-$400$ TeV ($E_\gamma \geq 4$ TeV), detected in XRECs, are mostly formed by primary $p$ with $E_0 = 1$-$10$ PeV, while γ-families with halo with $E_\gamma \geq 500$ TeV ($E\gamma \geq 4$ TeV) are mostly formed by primary p+He with $E_0 \geq 10$ PeV. The large number of γ-families observed in the PAMIR XRECs indicates a significant p+He fraction in the PCR spectrum up to $E_0 \sim 100$ PeV.

Thus, the halos observed in XRECs are mostly formed by p+He ($\geq 90\%$), the halo formation mechanism is described within the standard model of nuclear interactions, which allows using γ-families with halo to estimate the p+He fraction in PCR.

***Probabilities of forming γ-families with halo by PCR nuclei.*** The PCR mass composition in MC0 calculations is shown in Table 1. The calculations showed that 83% of all γ-families with halo are formed by primary $p$, 13% are formed by He nuclei, and $\leq 4\%$ are formed by heavy nuclei (Fig. 3). These calculations make it possible to estimate the p+He fraction in the PCR mass composition in the range $E_0 = 1$-$100$ PeV.

**Table 1.** PCR mass composition calculated by the MC0 model

| $E$, PeV | 1 | 10 | 100 |
|---|---|---|---|
| $p$,% | 33 | 26 | 20 |
| He,% | 22 | 17 | 15 |

The MC0 calculations show that in the case where PCR consist of only $p$, the number of observed halos should be 140 instead of 61 halos detected in the PAMIR XREC, 34 and 5 in the cases of pure He and heavy nuclei, respectively. It follows from the above that, taking into account the error (61 ± 8 of experimentally detected halos), the fraction of p+He and nuclei > He in the PCR mass composition at $E_0$ = 10 PeV should be ≥ 33%.

*Fraction of γ-families with multicenter halos.* In some cases, the halo consists of several centers. The fraction of multicenter halos is the parameter strongly depending on the PCR nucleus type (Table 2), i.e., the multicenter is the parameter sensitive to the p+He fraction in PCR.

It follows from Table 2 that γ-families with halo are mostly formed by primary *p* with a probably small He fraction. This conclusion is confirmed by the probabilities of forming γ-families with halo from PCR nuclei with $E_0 \geq 5$ PeV (Table 3).

**Table 2.** Fraction of γ-families with multicenter halos, formed by various nuclei

| P | He | C | Fe | MC0 | PAMIR |
|---|---|---|---|---|---|
| 0.25 ± 0.03 | 0.45 ± 0.09 | 0.59 ± 0.11 | 0.70 ± 0.03 | 0.28 ± 0.03 | 0.23 ± 0.07 |

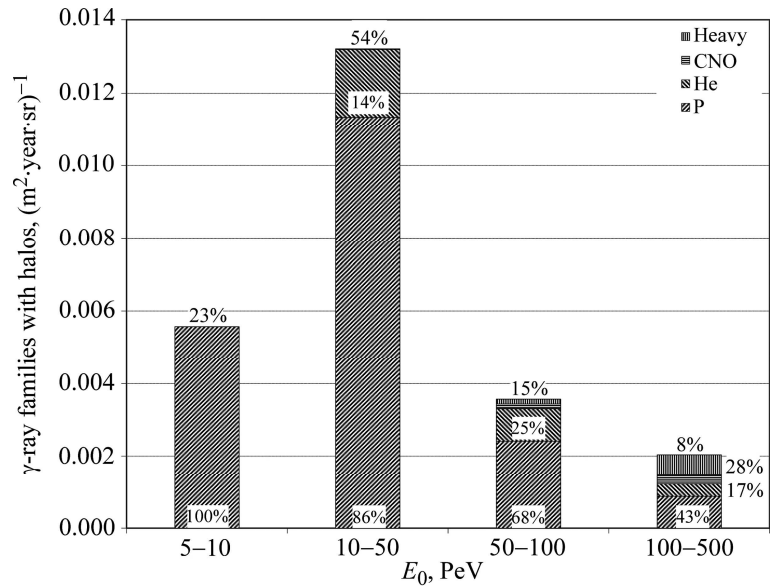

**Fig. 3.** Energy dependence of the flux of γ-families with halo and relative fractions of PCR nuclei forming these families.

**Table 3.** Halo formation probabilities

| p | He | > He |
|---|---|---|
| 1.76% | 0.44% | 0.13% |

Calculations show (Table 3) that the efficiency of the formation of γ-families with halo at $E_0 \geq 5$ PeV by primary *p* is four times higher than by He nuclei. This explains the preferential detection of γ-families with halo formed by *p* in XRECs.

*EAS age for estimating the PCR mass composition.* According to the data of the HADRON experiment [4] performed at the Tien-Shan at a height of 3330 m above the normal sea level (695 g/cm$^2$), 800000 EASs were analyzed, and the trend of the PCR mass composition to become heavier, from *p* to heavy nuclei in the range $E_0$ = 1-100 PeV (Fig. 4), was established.

The horizontal EAS development was determined using the NKG function,

$$\rho(N_e, r, S) = C(S) N_e (r/r_0)^{S-2} (1 + r/r_0)^{S-4.5},$$

where $\rho$ is the $e^{\pm}$ density at the distance $r$ from the EAS X axis, $N_e$ is the number of $e^{\pm}$ in EAS, $S$ is the EAS age parameter at the observation level, and $r_0$ is the Moliere radius.

The results of calculations of the dependence of the average EAS age $<S>$ on $N_e$ (the number of charged particles) for the level of the HADRON experiment in comparison with the EAS+XREC (HADRON) and EAS (KASCADE, Karlsruhe) experimental data, for which 1 million events were analyzed, are shown in Fig. 4. The EAS (KASCADE, Karlsruhe) data are reduced to the HADRON experiment level under the assumption of the linear dependence $S(N_e)$. The $S$ boundaries for p, He, C, and Fe, calculated for the HADRON, are shown in Fig. 4 by dashed lines: as $E_0$ increases, the EAS development maximum shifts to the atmosphere depth, and $S$ decreases.

Weighting of the average PCR mass composition in the energy range $E_0 = 5$-$50$ PeV corresponds to the HADRON data. However, according to HADRON data, the PCR mass composition remains mixed with a significant p+He fraction to $E_0 = 50$ PeV, whereas the KASCADE data show that the average EAS age at $E_0 = 50$ PeV corresponds to the dominant Fe fraction in the PCR mass composition.

An analysis of the parameter $S$, based on the HADRON data, shows that the PCR mass composition remains mixed to $E_0 = 100$ PeV.

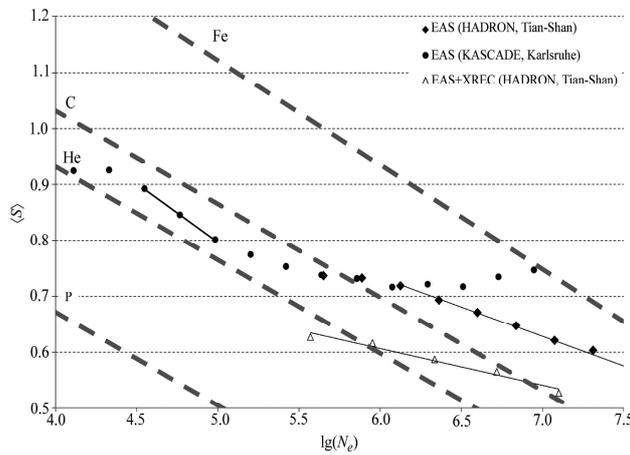
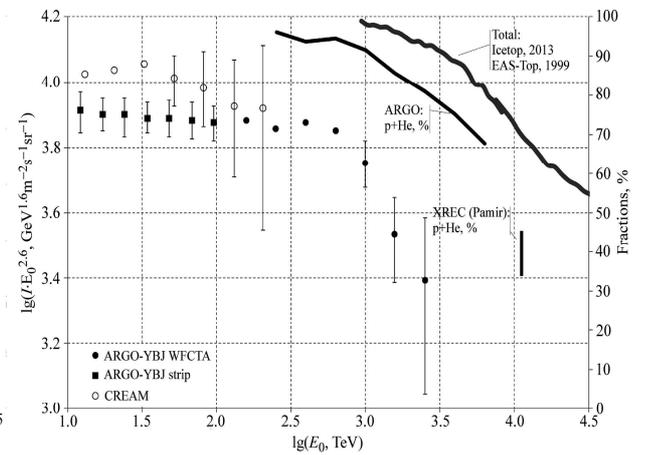

Fig. 4

Fig. 5

**Fig. 4.** Dependence of the EAS age $S$ on $N_e$. The dependences for EASs formed by primary p, C, He, and Fe (calculation by the MC0 for the HADRON) (dashed curves); HADRON data [4, 7] (diamonds and open triangles), and KASCADE [8] (closed circles).

**Fig. 5.** Dependences of the p+He flux on $E_0$ according to the ARGO and CREAM experimental data [5] (left axis) and the p+He fraction on $E_0$ according to the ARGO [6] and PAMIR XREC data (solid line, right axis).

***p+He fraction according to the ARGO, CREAM, and PAMIR XREC experimental data.*** The p+He fraction as a function of $E_0$ according to the ARGO, CREAM, and PAMIR XREC experimental data are shown in Fig. 5.

According to the ARGO data, the p+He fraction measured to $E_0 \sim 3$ PeV begins to rapidly decrease at $E_0 \sim 1$ PeV. The p+He fraction is shown in Fig. 5 by the solid curve [6]. At the same time, according to the PAMIR XREC data the p+He fraction at the same $E_0$ is significant and cannot be lower than 33%.

Similar results follow from an analysis of the EAS age according to the HADRON data. The PCR mass composition remains mixed with the trend to weighting at $E_0 = 1$-$100$ PeV.

***Conclusions.*** The calculations performed explained the nature of $\gamma$-families with halo within the standard model of nuclear interactions, which gives the good description of the PAMIR XREC experimental data.

(i) The halo phenomenon in γ-families is not an exotic event, but represents an overlap of individual dark spots on the *X*-ray film.

(ii) The estimation of the p+He fraction in the PCR mass composition at $E_0$ = 10 PeV, based on the analysis of the experimental number of γ-families with halo and the known probabilities of forming the halo by primary *p* and He, corresponds to ≥ 33%.

(iii) At $E_0$ = 1-100 PeV, the PCR mass composition remains mixed with the trend to weighting.